\documentclass[aps,prl,twocolumn,groupedaddress,showpacs]{revtex4}

\usepackage{graphicx}
\usepackage{amsmath}
\usepackage{amssymb}
\usepackage{bm}

\begin{document}
\title{Phase coherence of an atomic Mott insulator}
\author{Fabrice Gerbier, Artur Widera, Simon F{\"o}lling, Olaf Mandel, Tatjana Gericke and Immanuel Bloch}
\affiliation{Institut f{\"u}r Physik, Johannes
Gutenberg-Universit{\"a}t, 55099 Mainz, Germany.}
\date{\today}
\begin{abstract}
We investigate the phase coherence properties of ultracold Bose
gases in optical lattices, with special emphasis on the Mott
insulating phase. We show that phase coherence on short length
scales persists even deep in the insulating phase, preserving a
finite visibility of the interference pattern observed after free
expansion. This behavior can be attributed to a coherent admixture
of particle/hole pairs to the perfect Mott state for small but
finite tunneling. In addition, small but reproducible ``kinks''
are seen in the visibility, in a broad range of atom numbers. We
interpret them as signatures for density redistribution in the
shell structure of the trapped Mott insulator.
\end{abstract}
 \maketitle
%
%
%
A fundamental aspect of ultracold bosonic gases is their phase
coherence. The existence of long-range phase coherence, inherent
to the description of a Bose-Einstein condensate in terms of a
coherent matter wave, was experimentally demonstrated in
interferometric \cite{Andrews1997b,hagley1999b,bloch2000a} or
spectroscopic \cite{stenger1999b} experiments. More recently,
attention has been paid to fundamental mechanisms that may degrade
or even destroy long-range coherence, for example thermal phase
fluctuations in elongated condensates
\cite{petrov2000b,dettmer2001a,richard2003a,hellweg2003a}, or the
superfluid to Mott insulator (MI) transition undergone in optical
lattices \cite{jaksch1998a,greiner2002a,zwerger2003a}.

For a Bose-Einstein condensate released from an optical lattice,
the density distribution after expansion shows a sharp
interference pattern \cite{greiner2002a}. In a perfect Mott
Insulator, where atomic interactions pin the density to precisely
an integer number of atoms per site, phase coherence is completely
lost and no interference pattern is expected. The transition
between these two limiting cases happens continuously as the
lattice depth is increased. In the superfluid phase, a partial
loss of long range coherence due to an increased quantum depletion
has been observed for lattice depths below the MI transition
\cite{orzel2001a,hadzibabic2004a,schori2004a}. Conversely, in the
insulating phase, numerical simulations
\cite{kashurnikov2002a,roth2003a,schroll2004a} predict a residual
interference, although {\it long-range} coherence and
superfluidity have vanished.

In this Letter, we revisit this question of phase coherence
focusing on the insulating phase. We observe that the interference
pattern persists in the MI phase, and that its visibility decays
rather slowly with increasing lattice depth. We explain this
behavior as a manifestation of short-range coherence in the
insulating phase, fundamentally due to a coherent admixture of
particle/hole pairs to the ground state for large but finite
lattice depths. In addition, we also observe reproducible
``kinks'' in the visibility at well-defined lattice depths. We
interpret them as signature of density redistribution in the shell
structure of a MI in an inhomogeneous potential, when regions with
larger-than-unity filling form. Finally, the issue of adiabatic
loading in the lattice is briefly discussed.

\begin{figure}[hb!]
\includegraphics[width=8.6cm]{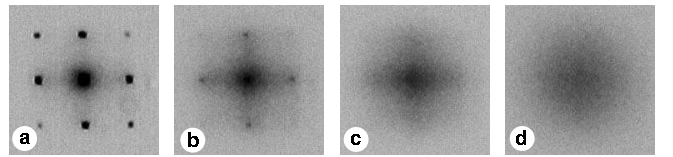}
\caption{Absorption images of an ultracold Bose gas released from
an optical lattice, for various lattice depths: {\bf a} 8 $E_{\rm
R}$, {\bf b} 14 $E_{\rm R}$, {\bf c} 18 $E_{\rm R}$, and {\bf d}
30 $E_{\rm R}$.} \label{images}
\end{figure}

In our experiment, a $^{87}$Rb Bose-Einstein condensate is loaded
into an optical lattice created by three orthogonal pairs of
counter-propagating laser beams (see \cite{greiner2002a} for more
details). The superposition of the lattice beams, derived from a
common source at a wavelength $\lambda_L=850~$nm, results in a
simple cubic periodic potential with a lattice spacing
$d=\lambda_L/2=425~$nm. The lattice depth $V_0$ is controlled by
the laser intensities, and is measured here in units of the
single-photon recoil energy, $E_{\rm R}=h^2/2m\lambda_L^2\approx
h\times3.2~$kHz, where $m$ is the atomic mass. The optical lattice
is ramped up in $160$ ms, using a smooth waveform that minimizes
sudden changes at both ends of the ramp. After switching off the
optical and magnetic potentials simultaneously and allowing for
typically $t=10-22~$ms of free expansion, standard absorption
imaging of the atom cloud yields a two-dimensional map of the
density distribution (integrated along the probe line of sight).

Four such images are shown in Fig.~\ref{images}a-d, for various
lattice depths. The density distribution of these expanding clouds
can be expressed as \cite{kashurnikov2002a,roth2003a,pedri2001a}
\begin{equation}\label{ntof}
n({\bf r})=\left( \frac{m}{\hbar t} \right)^3 \left|
\tilde{w}({\bf k}=\frac{m {\bf r}}{\hbar t})\right|^2
\mathcal{S}\left({\bf k}=\frac{m {\bf r}}{\hbar t}\right).
\end{equation}
In Eq.~(\ref{ntof}), the interference pattern is described by
\begin{equation}\label{Sk}
\mathcal{S}({\bf k})= \sum_{i,j} e^{i{\bf k}\cdot({\bf r}_i-{\bf
r}_j)}\langle \hat{a}_i^\dagger \hat{a}_j\rangle,
\end{equation}
where the operator $\hat{a}_i^\dagger$ creates an atom at site
$i$, and where $\tilde{w}$ is the Fourier transform of the Wannier
function $w({\bf r}_i)$. The Fourier relation (\ref{Sk}) shows
that long-range phase coherence, {\it i.e.} a correlation function
$\langle \hat{a}_i^\dagger \hat{a}_j\rangle$ slowly varying across
the lattice, is necessary to observe a sharp diffraction pattern
as in Fig.~\ref{images}a. However, above the MI transition
(Fig.~\ref{images}b, c and d), the interference peaks evolve into
a much broader, cross-like structure which weakens with increasing
lattice depth. This slow modulation corresponds to short-range
coherence, {\it i.e.} a correlation function $\langle
\hat{a}_i^\dagger \hat{a}_j\rangle$ whose range extends over a few
sites only.

\begin{figure}
\includegraphics[width=8.cm]{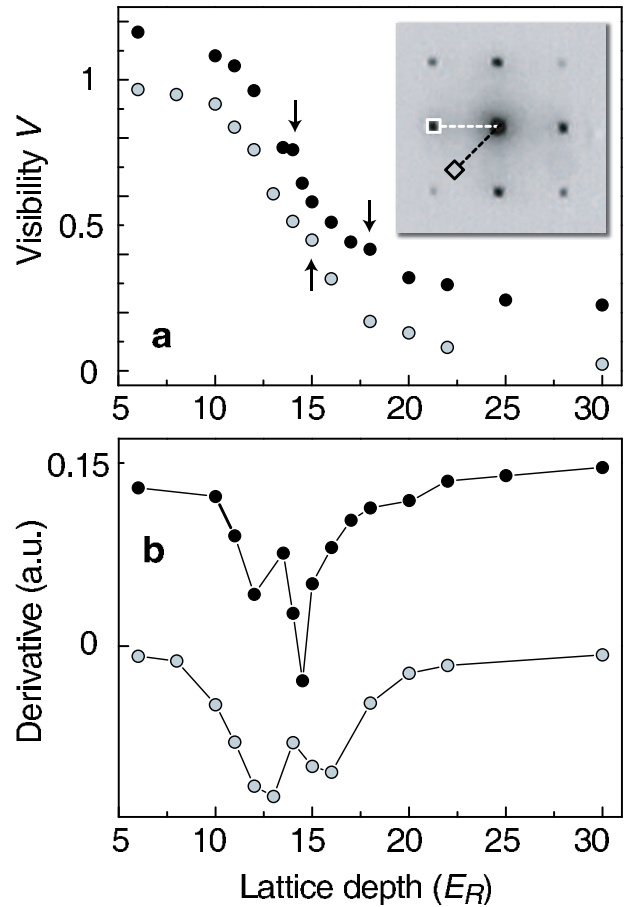}
\caption{{\bf (a)} Visibility of the interference pattern produced
by an ultracold cloud released from an optical lattice. The two
sets of data shown correspond to $3.6\times10^5$ atoms (grey
circles) and $5.9\times10^5$ atoms (black circles). The latter
curve has been offset vertically for clarity. Arrows mark
positions where ``kinks'' are visible. {\bf (b)} Numerical
derivative of the above curves.} \label{mainfig}
\end{figure}

To extract quantitative information from time-of-flight pictures
as shown in Fig.~\ref{images}, Eq.~(\ref{ntof}) suggests using the
usual definition of the visibility of interference fringes,
\begin{equation}\label{V}
\mathcal{V}=\frac{n_{\rm max}-n_{\rm min}}{n_{\rm max}+n_{\rm
min}}=\frac{\mathcal{S}_{\rm max}-\mathcal{S}_{\rm
min}}{\mathcal{S}_{\rm max}+\mathcal{S}_{\rm min}}.
\end{equation}
In this work, we measure the maximum density $n_{\rm max}$ at the
first lateral peaks of the interference pattern \cite{average},
({\it i.e.} at the center of the second Brillouin zone), whereas
the minimum density $n_{\rm min}$ is measured along a diagonal
with the same distance from the central peak (see inset in
Fig.~\ref{mainfig}a). In this way, the Wannier envelope is the
same for each term and cancels out in the division, yielding the
contrast of $\mathcal{S}$ alone (hence the second equality in
Eq.~(\ref{V})). Four pairs exist for a given absorption image, and
their values are averaged to yield the visibility. In previous
studies of the MI transition \cite{greiner2002a,schori2004a}, the
sharpness of the interference pattern was characterized by the
half-width of the central peak. Such a measure is possibly
sensitive to systematic effects, such as optical saturation and
mean field broadening. We expect our measure of contrast to be
much less sensitive to these effects, since it is calculated in
regions of the image where the density is lower.

We present here measurements of the visibility as a function of
lattice depth (typically in a range $6-30~E_{\rm R}$) at a given
total atom number. Each value was obtained as the visibility
averaged over approximately 10 independent images. Different atom
numbers (hence different filling factors) were investigated,
ranging from $6\times10^4$ to $6\times10^5$. Two illustrative sets
of data are shown in Fig.~\ref{mainfig}, corresponding to
approximately $5.9\times10^5$ atoms (black circles) and
$3.6\times10^5$ atoms (grey circles). For lattice depths larger
than $12.5~E_{\rm R}$, the system is in the insulating phase
\cite{greiner2002a}. Yet, the visibility remains finite well above
this point. For example, at a lattice depth of $15~E_{\rm R}$, the
contrast is still around $30 \%$, reducing to a few percent level
only for a rather high lattice depth of $30~E_{\rm R}$. We will
now show that such a slow loss in visibility is expected in the
ground state of the system.

As shown in \cite{jaksch1998a}, the physics of ultracold atoms in
an optical lattice can be described by the Bose-Hubbard
hamiltonian, given by the sum of a tunneling term,
$\mathcal{H}_{\rm t} =  - t \sum_{\langle i,j \rangle}
\hat{a}_i^\dagger \hat{a}_j $, plus an interaction term,
$\mathcal{H}_{\rm int}= \sum_i \frac{U}{2} \hat{n}_i \left(
\hat{n}_i-1 \right)$. Here $\hat{n}_i=\hat{a}_i^\dagger\hat{a}_i$
is the on-site number operator, $t$ is the tunneling matrix
element, the notation $\langle i,j \rangle$ restricts the sum to
nearest neighbors only, and $U$ is the on-site interaction energy
\cite{zwerger2003a}. In the experiments, an additional, slowly
varying potential $V_{\rm ext}({\bf r})$ is also present, and
favors the formation of a ``wedding cake'' structure of
alternating MI and superfluid shells
\cite{jaksch1998a,kashurnikov2002a,batrouni2002a}, which reflects
the characteristic lobes delimiting the MI phases in the phase
diagram of the Bose-Hubbard model \cite{zwerger2003a}.

\begin{figure}
\includegraphics[width=8.cm]{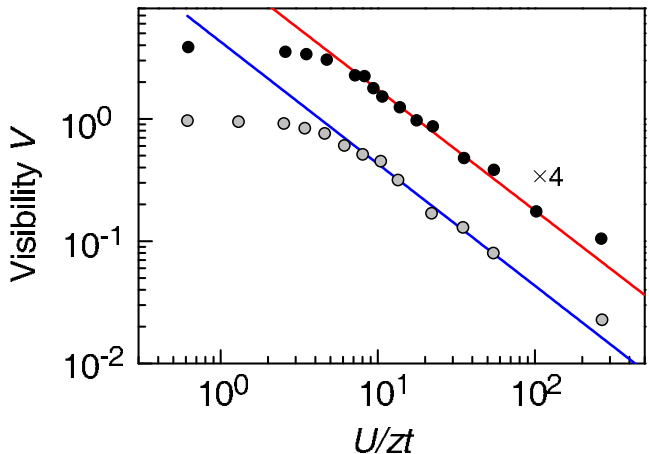}
\caption{Visibility of the interference pattern versus $U/zt$, the
characteristic ratio of interaction to kinetic energy. The data
are identical to those shown in Fig. \ref{mainfig}
($5.9\times10^5$, black circles, and $3.6\times10^5$ atoms, grey
circles). The former curve has been offset vertically for clarity.
The lines are fits to the data in the range $14-25~E_{\rm R}$,
assuming a power law behavior (see text).} \label{contrastvsUzt}
\end{figure}

To better understand the origin of a finite visibility, we
consider a homogeneous system with filling factor $n_0$. In the
limit of infinitely strong repulsion, $U/t\rightarrow\infty$, the
ground state is what we call a ``perfect'' Mott insulator, {\it
i.e.} a uniform array of Fock states, $|\Psi \rangle_{\rm MI} =
\prod_i|n_0 \rangle_i$. This corresponds to a uniform
$\mathcal{S}=n_0$ and zero visibility. To a good approximation,
the actual ground state for a finite ratio $U/t$ can be calculated
by considering the tunneling term as a perturbation to the
interaction term. To first order in $t/U$, this yields
\begin{equation}
|\Psi^{(1)} \rangle \approx |\Psi \rangle_{\rm MI} + \frac{t}{U}
\sum_{\langle i,j\rangle} \hat{a}_i^\dagger \hat{a}_j |\Psi
\rangle_{\rm MI}.
\end{equation}
The ground state thus acquires a small admixture of
``particle-hole'' pairs ({\it i.e.} an additional particle at one
lattice site and a missing one in a neighboring site), which
restores short-range coherence and a corresponding weak modulation
in the momentum distribution, $\mathcal{S}({\bf k}) \propto n_0 -2
n_0(n_0+1)t({\bf k})/U$, where $t({\bf k})=-2t \sum_{\nu=x,y,z}
\cos(k_\nu d)$ is the tight-binding dispersion relation. The
corresponding 2D visibility (integrated along one direction) is
\begin{equation}\label{Vcalc}
\mathcal{V} \approx \frac{4}{3}(n_0+1) \frac{z t}{U}.
\end{equation}
In Eq.~(\ref{Vcalc}), $z=6$ is the number of nearest neighbors in
a 3D cubic lattice.

\begin{figure}
\includegraphics[width=8.cm]{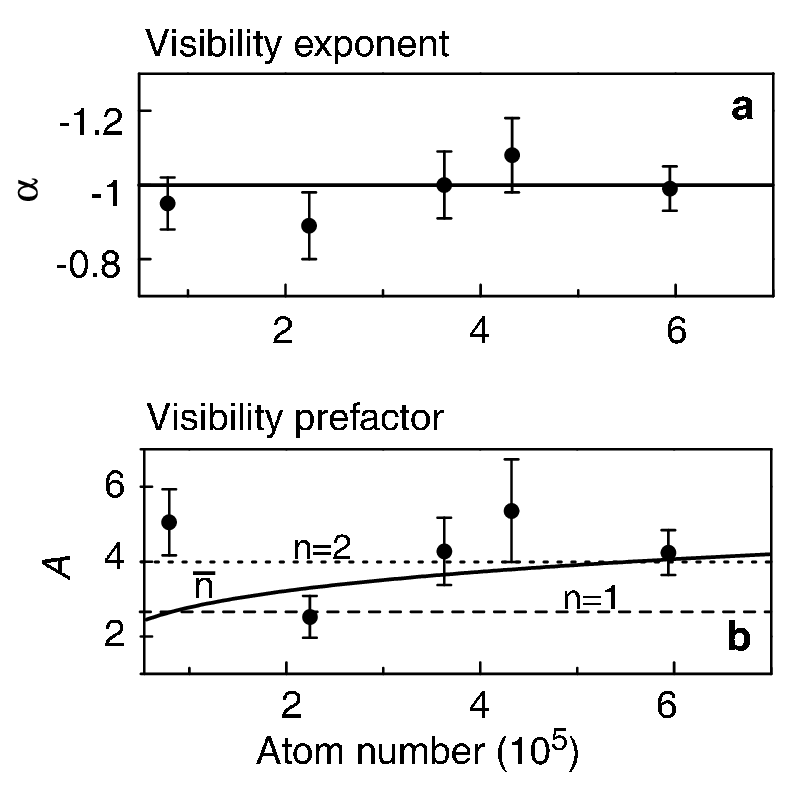}
\caption{Exponent $\alpha$ ({\bf a}) and prefactor $A$ ({\bf b})
extracted from a power law fit $A (U/z t)^{\alpha}$ to the
visibility data in Fig. \ref{contrastvsUzt}, plotted versus total
atom number. The solid line indicates the expected exponent
$\alpha=-1$. In ({\bf b}), we also indicate the prefactor expected
for uniform MI with filling factor $n_0=1$ (dashed line) and
$n_0=2$ (dotted line), as well as an extrapolation for the average
filling calculated at a lattice depth of $30~E_{\rm R}$ (solid
line).} \label{fit_contrast}
\end{figure}

To compare with the experiment, we show in
Fig.~\ref{contrastvsUzt}a the visibility against $U/zt$ in a
log-log plot. For lattice depths $V_0 \geq 14~E_{\rm R}$
(corresponding to $U/zt\geq 8$), the data matches the inverse law
expected from Eq.~(\ref{Vcalc}). This has been verified by fitting
the data in this range to a general power law $A (U/zt)^{\alpha}$
(solid lines in Fig.~\ref{contrastvsUzt}). We obtain an average
exponent $\alpha=-0.98(7)$ in agreement with the prediction (see
Fig.~\ref{fit_contrast}a). In Fig.~\ref{fit_contrast}b, the fitted
prefactor is plotted as a function of atom number. Inspired by Eq.
(\ref{Vcalc}), we compare it to $4(\overline{n}+1)/3$, where
$\overline{n}$ is the average filling factor calculated at a
lattice depth of $30~E_{\rm R}$ using a mean-field approximation
\cite{sheshadri1993a,vanoosten2000a}. We find that this
extrapolation of Eq.~(\ref{Vcalc}) to our trapped system indeed
yields the correct order of magnitude (see
Fig.~\ref{fit_contrast}b). We thus consider the agreement between
our experimental results and the simple relations derived above as
a conclusive evidence for the presence of particle-hole pairs,
characteristic of the ground state of the Bose-Hubbard
hamiltonian.

In addition to the smooth decay discussed above, the visibility
shows small ``kinks'' at specific lattice depths (indicated by
arrows in Fig.~\ref{mainfig}a). They are systematically observed
in our data, and their positions are reproducible. In the
derivative plot (Fig.~\ref{mainfig}b), they appear as narrow
maxima on a smoother background. We obtained the kink positions by
taking the middle point between two adjacent gaussian peaks with
negative amplitudes fitted to the data. The most prominent kink
occurs on average for a lattice depth of $14.1(8)~E_{\rm R}$, with
a statistical error indicated between parentheses. For the largest
atom numbers ($4.2\times 10^5$ and $6\times 10^5$), a similar but
much weaker kink is also visible around 16.6(9) $E_{\rm R}$ (see
upper curves in Fig.~\ref{mainfig}). These values are close to
14.7 $E_{\rm R}$ and 15.9 $E_{\rm R}$, the lattice depths where MI
regions with filling factor $n_0=2$ or $3$ are respectively
expected to form for our parameters \cite{vanoosten2000a}. We thus
propose that the observed kinks are linked to a redistribution in
the density as the superfluid shells transform into MI regions
with several atoms per site. We were recently informed that
similar features were reproduced numerically for one-dimensional
trapped systems with a small number of particles \cite{roth2005}.

We have considered the dependence of the visibility on the time
over which the optical lattice was ramped from zero to its final
value, for a specific lattice depth of $V_0=10~E_{\rm R}$. The
visibility was considerably degraded for the shortest ramp time of
$20~$ms, but reached a ramp-independent value for ramp times
larger than $T_{\rm ad}\sim100~$ms (to be compared to the $160~$ms
time used in visibility experiments). We note that $T_{\rm ad}$
for this lattice depth of $V_0=10~E_{\rm R}$ is significantly
longer than the microscopic time scales of the system, such as the
tunneling time or the trapping periods. We note also that at the
largest lattice depth we use here ($V_0=30~E_{\rm R}$), the
observed visibility is systematically above the power law fit in
Fig.~\ref{contrastvsUzt}, indicating a breakdown of adiabaticity.
By comparing the data to the fitted curve, we expect this to occur
for $V_0\approx29~E_{\rm R}$ ($U/zt\approx200$), which agrees with
the calculated depth of $32~E_{\rm R}$ for which the ramping time
$160~$ms becomes smaller than the calculated tunneling time $h/z
t$.

Although a complete study is beyond the scope of this Letter,
these observations suggest that different dynamical processes are
involved in the loading, depending on whether the gas is in the
superfluid or in the MI phase. In the superfluid phase, the ramp
time has to be slow enough not to excite long-lived collective
excitations. In the MI phase, these excitations acquire an energy
gap, which makes single particle tunneling the dominant dynamical
process. In this case, the final tunneling time increases with
final lattice depth, and eventually becomes so long that the
system basically freezes out at some lattice depth, estimated here
to be $29~E_{\rm R}$.

In conclusion, we have studied the visibility of the interference
pattern produced by an ultracold Bose gas released from a deep
optical lattice. A non-vanishing visibility in the MI phase is
observed and explained by the coherent admixture of particle-hole
pairs to the insulating ground state, which preserves local phase
coherence. This intrinsic limitation to the ``quality'' of a MI
has important implications for various quantum information
processing schemes, where the MI plays a central role
\cite{rabl2003a,pupillo2004a,demarco2005a}. In addition, we
observe small but reproducible kinks in the visibility curve. We
interpret them as the signature of density redistribution in the
shell structure of the cloud as MI with several atoms per site are
expected to form. Finally, a recent paper \cite{gangardt2004a}
suggests that in a planar array of one-dimensional Bose gases, the
visibility might be further reduced when correlations build up in
each tube, {\it i.e.} upon entering the Tonks-Girardeau regime.
Experimental study of these effects seems within reach with the
methods presented in this paper.

We would like to thank Dries van Oosten, Paolo Pedri and Luis
Santos for useful discussions. Our work is supported by the
Deutsche Forschungsgemeinschaft (SPP1116), AFOSR and the European
Union under a Marie-Curie Excellence grant. FG acknowledges
support from a Marie-Curie Fellowship of the European Union.
%
%

\begin{thebibliography}{28}
\bibitem{average} The value is obtained by averaging over
$3\times3$ pixels around each position. We have checked that the
visibility obtained this way was almost independent of the size of
the integration region used to find $n_{\rm max}$ and $n_{\rm
min}$.
\end{thebibliography}


\end{document}